# Design and Optimization of Gregorian-based Reflector Systems for THz Imaging System Optics

Mohammad Hossein Koohi Ghamsari
*Department of Electrical Engineering*
Sharif University of Technology
Tehran, Iran
mohammadhossein.kgh@ee.sharif.edu

Mehdi Ahmadi-boroujeni
*Department of Electrical Engineering*
Sharif University of Technology
Tehran, Iran
ahmadi@sharif.edu

Saeed Babanejad
*Department of Physics*
University of Tabriz
Tehran, Iran
saeed.b95@chmail.ir

*Abstract*— Application of reflector antenna configurations in designing the optical section of terahertz imaging systems has attracted great attention during recent years. Advantages of reflective optics in comparison with refractive optics is one of the main reasons for this trend. Specular reflection in broad spectral range, no or little reflection and transmission loss, no chromatic aberration, high reliability in high power applications and even cancellation of off-axis aberrations as a result of precise optimized configuration are just some of these advantages. Moreover, utilizing novel satellite communication reflector antenna systems in an optical section of imaging systems has improved many imaging performance factors and provided a reliable platform for different imaging objectives. In this paper, we will design a Gregorian-based reflector antenna system for the optical section of terahertz imaging systems and using optical techniques, we optimize the structure for optimum imaging performance.

## I. Introduction

Extensive use of terahertz frequency radiation in science and technology has increased rapidly over the recent decades. Today, applications of THz band have expanded from communication systems, biomedical and security imaging systems, THz time-domain spectroscopy of semiconductors and semiconductor nanostructures, Non-destructive testing and molecular spectroscopy, THz microscopy to the area of astronomy and astrophysics [1].

Regarding these characteristics, using THz in defense and security imaging systems has attracted more attention in recent years. More specifically, using THz in standoff passive and active imaging systems which provides acceptable resolution images of the body at standoff distances without any exposure to ionizing radiation has been one of the important and growing research areas in imaging systems.

Several types of research and commercial developments for passive thermal video-rate imaging systems have been reported. For example, [1] reports a standoff passive technology terahertz camera at an 8-meter distance which uses the bolometers in the detector section. However, important disadvantages of passive approaches like low contrast and image resolution, as well as relatively attenuated and distorted received signals as a result of transmission from thick layers of clothing have prevented its use in many high reliability demanding applications. Active imaging systems solve some of these challenges with the efficient use of different illumination techniques with appropriate transmitting power and frequency depending on desired imaging objective. High contrast especially at high frequencies, penetration through thick clothing, and achieving a very high signal-to-noise ratio (SNR) as a result of no fundamental detection sensitivity limit are advantages of active imaging techniques.

Utilizing reflector antenna configurations in active and passive imaging systems has attracted great research interest because of the advantages of reflective optics against refractive optics. In the frequency-modulated based imaging system described in [2], a folded-path antenna is used with imaging radar optics to focus the beam at a 4 m standoff range. Low scanning speed as a result of the mechanical rotation of the entire radar platform is one of the main drawbacks of this work. [3] and [4] proposed a separate station for rotation to increase the imaging time in pixel-by-pixel scanning radar-based active imaging systems. In these works, a separate special lens is used for collimating the reflected beam. In [5] a different scanning strategy is proposed in which two-scanning element is used for fast and efficient scanning and focusing of the target. In [6] two reflectors in an off-axis rotating configuration are implemented for scanning in the target plane. Moreover, [7] proposed a real-time terahertz camera and a periscope-based reflective conical scanner that sweeps out the target plane at a standoff range of 8m. A 670 GHz Gregorian reflector system radar imaging optics for a standoff distance of 25 m implemented in [8]. [9] reported a noticeable improvement in the scanning rate and standoff distance of terahertz imaging utilizing previous proposed confocal Gregorian reflector antennas configuration. [10] achieved 3-D images with centimeter cell resolution at a distance of 8 m with the frequency of 300 GHz. Synthetic aperture focusing is another useful imaging technique that is reported to apply to systems such as [11].

The paper is organized as follows. In Section II, the geometry and design of a parabolic Gregorian reflector system are presented. Section III reviews the equivalent theory of paraboloidal and ellipsoidal reflectors and in Section IV we extend the parabolic-based Gregorian design to an off-axis ellipsoidal Gregorian system. We show that optimizing this off-axis structure will result in a near-ideal focusing spot in the desired standoff distance.

## II. CONFOCAL PARABOLOIDAL GREGORIAN REFLECTOR SYSTEM

Gregorian reflector antenna structure is a popular configuration in satellite communications and optical telescopes. We consider first a classical on-axis symmetric Gregorian telescopic antenna configuration with a paraboloidal main reflector and an elliptical reflector as subreflector. Figure 1 (a) shows a typical configuration and figure 1 (b) is a design example and ray tracing pattern on this structure which is calculated and optimized using MATLAB software.

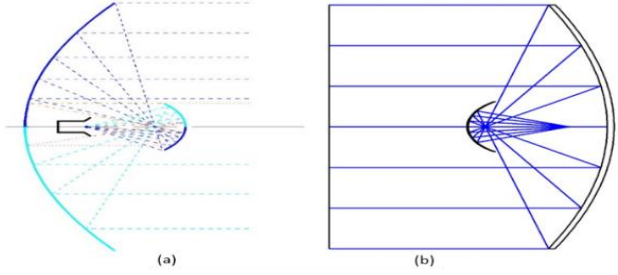

Fig. 1. Classical on-axis Gregorian telescopic reflector structure: (a) Geometric structure [21]. (b) Design example.

The idea behind this structure, which we will also use in the next section for designing confocal ellipsoidal Gregorian, is that the elliptical subreflector has two foci. Feed antenna (or detector in the receiver section of imaging system) is located at the first foci and the second foci is collocated with the main foci of the parabolic reflector (common focus). A modern extended version of classical Gregorian reflector structure is confocal paraboloidal Gregorian reflector system (CPGRS) as shown in Fig. 2. This structure uses four reflectors (three parabolic reflectors and one flat reflector or mirror) which except the main reflector, three other reflectors are rotated to be aligned with respect to geometrical optics rules. Similar to classical Gregorian structure, the main section of CPGRS consists of a main reflector and a subreflector sharing a common focus, except that in this structure both are parabolic surfaces of revolution [12]-[15].

Based on the literature, confocal Gregorian proved superior scanning performance in comparison with other similar structures [15], [16]. It has been proved that by proper use of reflectors geometry and precise alignment of components in a CGRS, off-axis aberrations, astigmatism and coma, can be cancelled and a quadratic path length error can be achieved. As a result, CGRS has superior scanning performances in comparison with other reflector antenna systems [15].

A design example of CPGRS with associated 2D and 3D ray tracing patterns is shown in Figure 3 (a) and figure 3 (b), respectively. In this reflector system, a collimated beam (from far field) is concentrated to the common focus of main reflector and subreflector. After focusing, the parabolic subreflector again collimate the beam toward the scanning mirror. Finally, the rotated beam will be focused by a feed reflector in the position of the transmitting antenna or receiving detector. If this configuration is intended to transmit and receive the beam simultaneously, which is the case in our design, a beamsplitter is mandatory in the feed section which we deliberately omitted in the schematic for simplifying the ray tracing pattern diagrams.

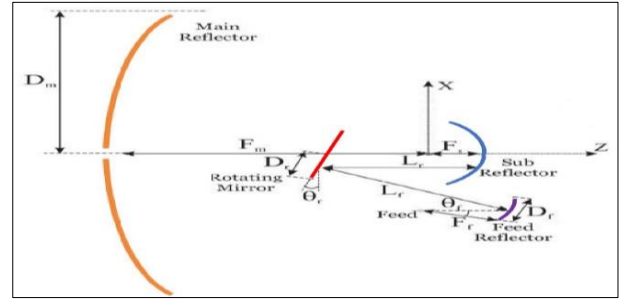

Fig. 2. Classical CPGRS geometry [8].

As it is evident, a good focusing point is achieved in the feeding (or detecting) point. Furthermore, we can calculate optical diagnostic curves to have a better performance criterion for designed optics. As we can see in Fig. 4, the RMS spot radius in image plane is calculated about 5mm with respect to chief ray. Although because of subreflector and scanning mirror blockage and consequent nontrivial magnitude of aberrations in the designed optical structure, an ideal minimum spot radius and focusing in the transmitter/receiver point is not achieved, we can still have an acceptable spot radius in the target point which is sufficient for many applications. The first order distortions of designed optics can be evaluated by calculating the optical path difference (OPD) error for the main reflector area [17]. Figure 4 represents optical path length error as a result of ray tracing. This figure shows the difference between chief rays and marginal rays in a standard ray tracing computation.

Optical or modulation transfer function (MTF) is defined as the Fourier transform of point spread function (PSF). It can also be interpreted as the auto-correlation of the entrance pupil function of the entire optics. If we define line spread function (LSF) as a 1-D PSF on x-axis, we have [18]

$$LSF(x) = \int_{-\infty}^{+\infty} PSF(x,y)\, dy \quad (1)$$

Therefore, 1-D MTF can be calculated from the LSF as

$$MTF(f_x) = |\mathcal{F}[LSF(x)]| = \left| \int_{x_{min}}^{x_{max}} LSF(x) e^{i2\pi x f_x} dx \right| \quad (2)$$

Where $f_x$ is the spatial domain variable of x. Analytical computation of MTF can confront difficulties in some cases. Therefore, in a numerical sense, 1-D MTF can also be computed by taking discrete Fourier transform (DFT) from the LSF. Following this approach and assuming $n^{th}$ pixel position as $y_n$, we can calculate MTF as

$$\text{MTF} = \sum_{n=0}^{N-1} y_n \left[ \cos\left(k\frac{2\pi}{N}n\right) - i\sin\left(k\frac{2\pi}{N}n\right) \right] \quad (3)$$

For $k \epsilon [0, N-1]$. Calculated polychromatic diffraction MTF against spatial frequency is shown in Figure 5 (c). Calculated polychromatic diffraction MTF in Figure 5 (c) can be assumed as a merit function for evaluating the ability of designed imaging optics to transfer spatial contrast details of object points into image points. Comparing the results of this figure, which is plotted against spatial frequency, with common standard MTF plots of imaging systems, shows a relatively good transfer contrast from the object to the image at a particular resolution.

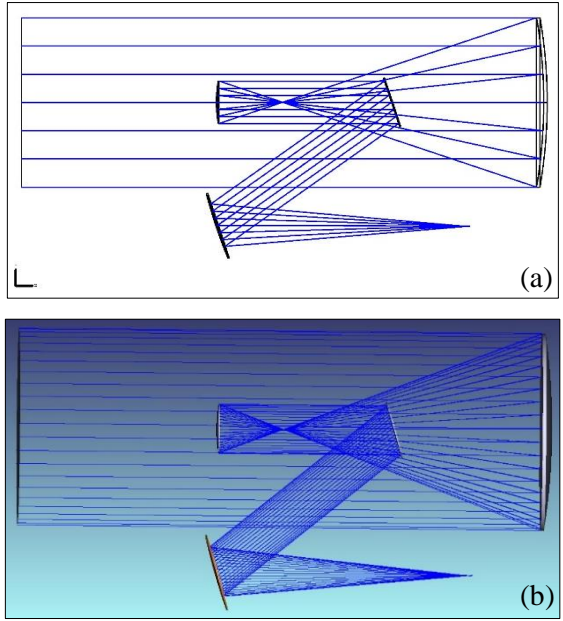

Fig. 3. Design example of on-axis CPGRS: (a) 2D ray tracing. (b) 3D ray tracing.

It is also useful to observe that if we additionally rotate tertiary feed-reflector, we can achieve an ideal focusing point, as shown in figure (6). Although, considering the new position of the transmitter/detector, this near aberration-free focusing point is achieved again at the cost of more ray blockage in the optical structure.

### III. REFLECTORS EQUIVALENCE THEORY

In the previous section, structure of a classical CPGRS has been studied. The parabolic main reflector in this structure is used to focus in the far field. However, usually in THz imaging systems targets are placed in the Fresnel region of the imaging antenna.

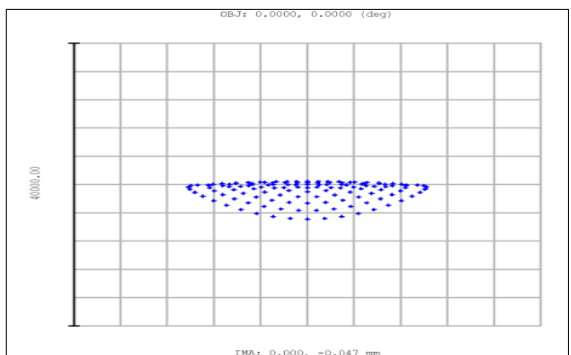

Fig. 4. Image plane Spot diagram of CPGRS.

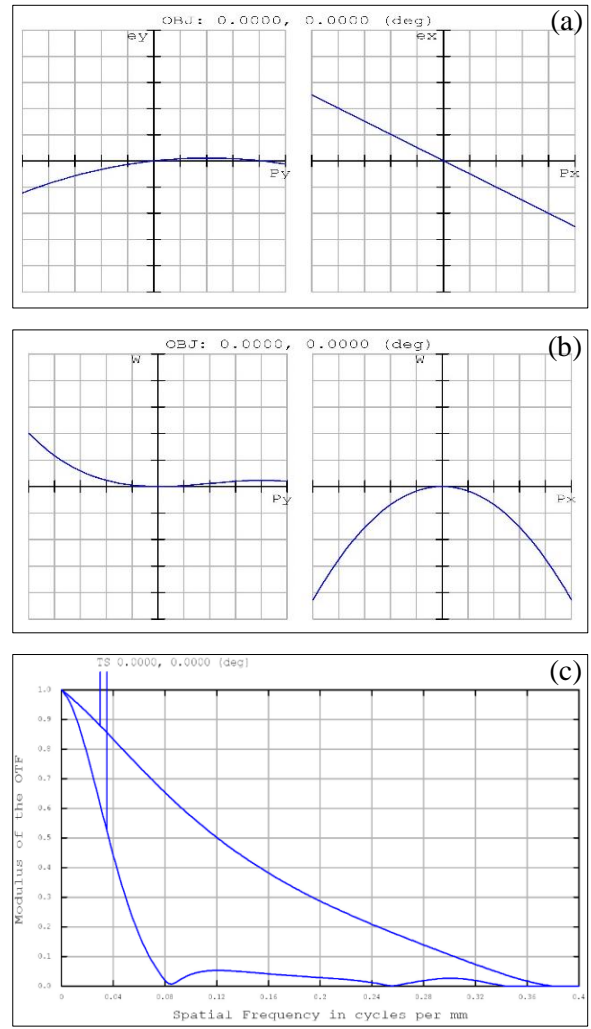

Fig. 5. Optical curves of designed on-axis CPGRS: (a) Transverse ray fan plot. (b) Optical Path Difference. (c) Polychromatic Diffraction MTF.

Therefore, in order to have near field focusing and diffraction limited resolution of the target objects, parabolic main reflector must be replaced with an ellipsoidal reflector surface. This theory can be efficiently used for designing antennas with specific near field characteristics using far field design techniques [8]. For example, reference [19] used this optical approximation theory to compensate for cross-polarization of the antenna.

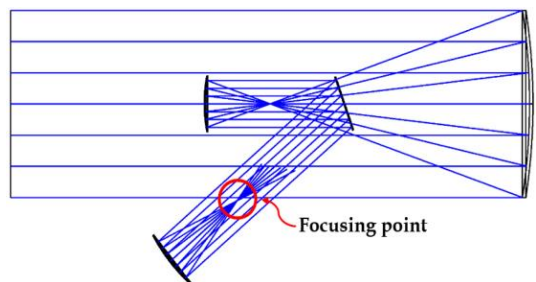

Fig. 6. 2D ray tracing of rotated tertiary feed-reflector.

Fig. 7 shows a design example of this equivalence theory which we will use in the next section for designing CEGRS

configuration. In figure 7 (a) we designed a parabolic surface for a focal length of 1m and in figure 7 (b), we derived an equivalent ellipsoidal form of this surface with main and second focal points in 1m and 25m, respectively.

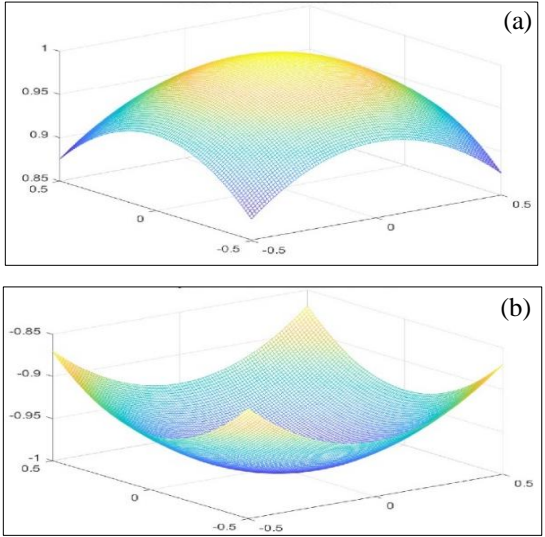

Fig. 7. Optical equivalent surfaces: (a) Parabolic reflector. (b) Equivalent ellipsoidal reflector.

## IV. OFF-AXIS CONFOCAL ELLIPSOIDAL GREGORIAN REFLECTOR SYSTEM

In this section, based on the derived equivalent ellipse surface in the previous section and using offset and off-axis optical techniques, we design and optimize an off-axis Confocal Ellipsoidal Gregorian Reflector System (CEGRS) for focusing in the near field. First, we consider the design of a popular off-axis variation of classical Gregorian structure known as unobscured Gregorian, as shown in Fig. 8. This confocal version of Gregorian utilizes a main parabolic reflector in combination with an ellipsoidal subreflector to achieve an ideal direct blockage-free focusing system. Three rays from different field angles are shown in this figure for example. With the use of idea behind this simple off-axis structure and manipulation of previous designs, we get the final optimum result of Fig. 9. Note that unlike classical design, in this configuration parabolic feed reflector is not rotated. Furthermore, the ellipsoidal surface is shaped so it can focus in standoff distance of 25 m which object is located. As it is evident in figures 9 (a)-(d), by proper selection of reflector types and precise alignment (position and tilting) of reflectors, one can achieve relatively ideal reflective optics with respect to aberrations.

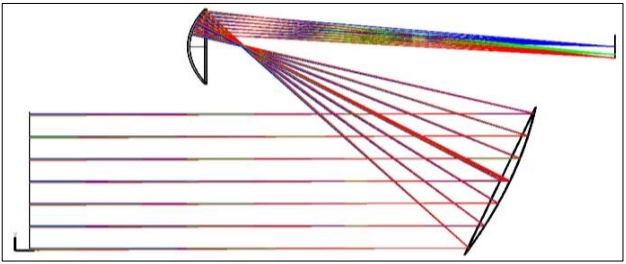

Fig. 8. Design example for an off-axis unobscured Gregorian.

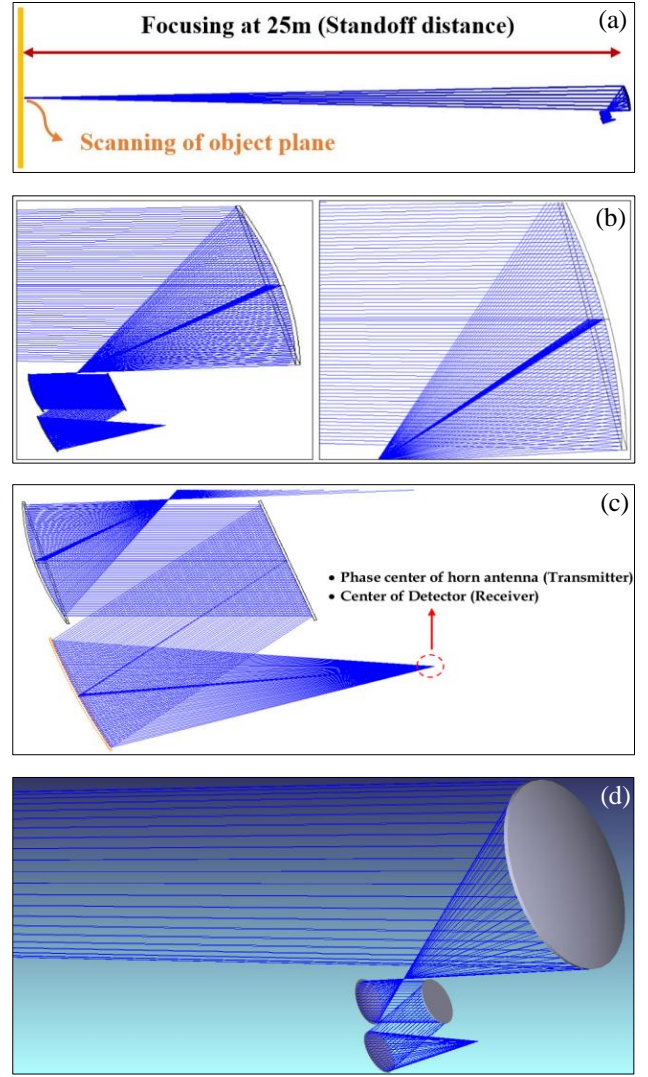

Fig. 9. Final optimized CEGRS: (a) 2D ray tracing. (b) and (c) zoom-in on different parts of the structure. (d) 3D ray tracing.

In the next step, RMS spot radius with respect to centroid is selected as the optimization goal and a Gaussian Quadrature pupil integration method is utilized. Figures 10 (a)-(d) show the final results of designed and optimized imaging optics. we can quantify the impulse response of our designed optics using point spread function (PSF) which can be derived from diffraction integrals. As shown in Figures 10 (a) and (b), a relatively sharp spot-like PSF is calculated for designed optics which, when convolved with the object function, can extract a high-quality image. Moreover, Figures 10 (c) and (d) show FFT and Huygens PSF. Note that FFT PSF is calculated in pupil space and Huygens PSF is calculated in image space. Comparing these figures with conventional standard optical figures, again predict a high-quality imaging performance for the designed optics. As we can see, using an off-axis technique, at a cost of more complex optical alignment and higher manufacturing costs, yields a better optical performance as a result of removing ray obstructions that exist in on-axis designs.

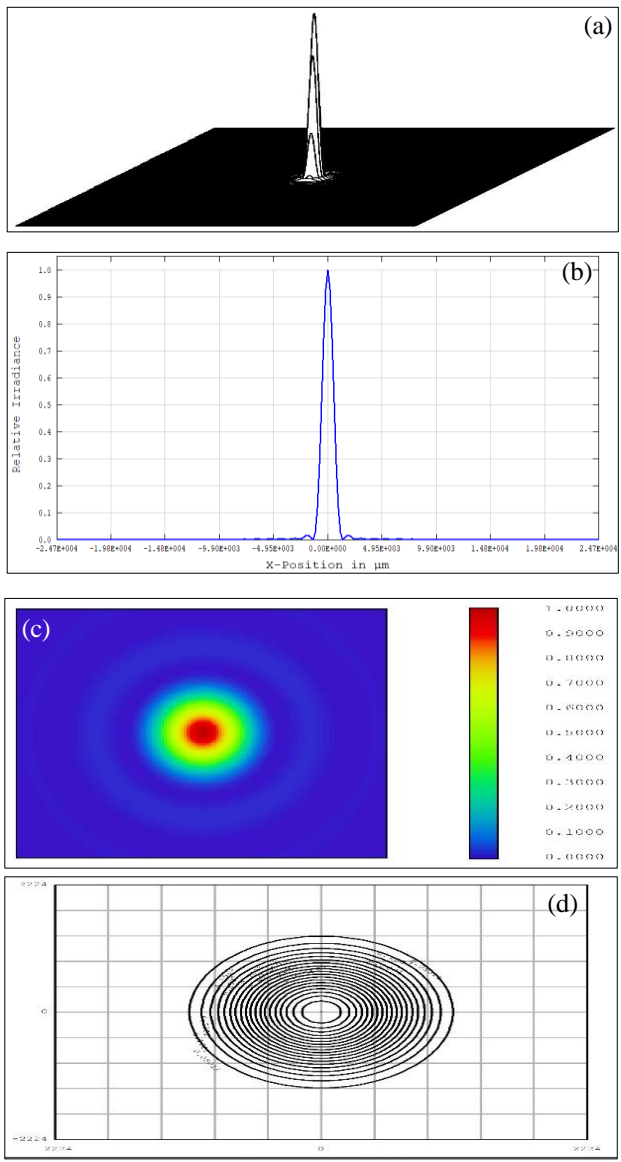

Fig. 10. PSF of designed CEGRS optics: (a) 3D polychromatic FFT PSF. (b) FFT PSF cross section. (c) Huygens PSF. (d) Huygens PSF contour.

## CONCLUSION

In this paper, we studied Gregorian-based reflector antenna systems for optics of modern THz imaging systems. By applying paraboloidal/ellipsoidal design equivalence theory to CPGRS configuration, an optimized near field focusing version of CPGRS was presented. Ray tracing diagrams and numerical simulations were performed and performance criterion parameters and plots were investigated, showing acceptable performance of the designed optics for THz imaging systems.